**THREE KINDS OF AI ETHICS**

Emanuele Ratti[1]

Department of Philosophy, University of Bristol

**Abstract**. There is an overwhelming abundance of works in AI Ethics. This growth is chaotic because of its volume, its multidisciplinary nature, and how sudden it has been. This makes difficult to keep track of debates, and to systematically characterize goals, research questions, methods, and expertise required by AI ethicists. In this article, I show that the relation between 'AI' and 'ethics' can be characterized in at least three ways, which correspond to three well-represented kinds of AI ethics: ethics and AI; ethics in AI; ethics of AI. I elucidate the features of these three kinds of AI Ethics, characterize their research questions, and identify the expertise that each kind needs. I also show how certain criticisms to AI ethics are misplaced, as being done from the point of view of one kind of AI ethics, to another kind with different goals. All in all, this work sheds light on the nature of AI ethics, and sets the groundwork for more informed discussions about the scope, methods, and training of AI ethicists.

## 1. INTRODUCTION

Literature in AI Ethics has recently exploded. What characterizes this growth is not just how quick it was, but also how varied it is from a disciplinary perspective. Contributions to AI Ethics come from a number of directions, including philosophy, human-computer interaction, political theory, computer science, social sciences, and law, just to name a few. It is difficult to keep track of AI ethics trends, because key concepts (e.g., algorithmic bias; surveillance; explainability, trustworthy AI; AI safety) come and go pretty quickly. Because of this sudden and chaotic growth, AI Ethics has not yet had an opportunity to stabilize as a unique discipline with its own specific questions, methodologies, exemplars, and good practices. Given these unclear boundaries, lack of well-characterized expertise needed, and vague research questions, there is much skepticism over AI ethics. This skepticism has especially emerged in criticisms revolving around AI ethics' applicability and lack of reflexivity. The former refers to the lack of guidance and methods for implementing ethics into the daily activities of AI practitioners (Morley et al 2020), while the second refers to the idea that, when the former challenge is overcome, AI practitioners seem to uncritically assume off-the-shelf tools for the ethical dimension of their work (Fazelpour and Danks 2021), and keep normative reflection to a bare minimum. To formulate these issues differently, there are three specific questions concerning AI ethics as a discipline that are capturing the attention:

1. What kind of questions does AI ethics aim to answer?
2. What kind of expertise do AI ethicists need?
3. Are current criticisms to AI ethics warranted?

This difficult situation faced by AI ethics is unlike much older applied ethics disciplines like clinical ethics or research ethics. These have their own standardized trainings, professional figures, and even textbooks (Beauchamp and Childress 2009; Shamoo and Resnik 2015). AI Ethics is at the very beginning of such a process of standardization. For instance, there is the *AAIE*[2], which is an association created to promote the professional interests and development of AI ethicists around the world. There are occasionally summer schools aimed at providing crash courses on AI Ethics and its many facets.

[1] mnl.ratti@gmail.com

[2] https://ethicists.ai/

But what is missing though is an attempt to systematize AI Ethics as a discipline, in such a way that questions 1 and 2 can be answered, and 3 addressed.

The goal of this article is to provide a new conceptualization of AI ethics that can address the three questions above. The starting point is to consider that AI Ethics lies at the conjunction of the terms 'ethics' and 'AI', and here I show how relationships between these two words can be conceived in rather different ways. In particular, I characterize three specific kinds of AI Ethics: Ethics *and* AI, Ethics *in* AI, and Ethics *of* AI. The benefits of this characterization are not merely taxonomical. Instead, what I will show is that these three kinds of AI Ethics presuppose different relations between the two terms 'AI' and 'ethics', and they ask different questions about the relevance of ethics to AI and *vice versa*. As a consequence, the goals of AI Ethicists will be rather different across these three kinds of AI Ethics, and their methodologies and scope will be as well. Different trainings will be required as a result of the kind of AI ethics that an institution or a company is interested in. In this way, questions 1 and 2 are answered comprehensively. But there is also another notable consequence, which will address question 3. Classic criticisms raised against the status of AI Ethics are often misplaced: they are raised from the point of view of the goals of one kind of AI Ethics, to another kind of AI Ethics that does not have the same goal.

In order to show the different relations between 'AI' and 'ethics' underpinnings the three kinds of AI Ethics, I will rely on the Capability Approach (Sen 1999; Nussbaum 2011) as an illustrative example through which elucidating the general characteristics of each kind of AI Ethics. This is especially useful to show how the same normative framework can serve three different goals within AI Ethics, and as a result how such goals require three different kinds of skillsets and expertise. This illustrative example will be complemented with concrete examples from recent scholarship in AI Ethics. While I have not followed any specific methodology for identifying the relevant literature, Supplementary Table 1 provides an exemplificative list of papers in AI ethics, and how they can be classified in light of my account. The table is not final, and it can be found online as a regularly updated document, where new papers or papers read in the past are added[3].

The structure of the article is as follows. First, there are preliminary considerations to lay out. In Section 1.1, I will give a succinct description of the Capability Approach, and in 1.2 I will define important terms that will be used in the article. In Section 2, I will illustrate themes and problems from Ethics *and* AI. In Section 3 and 4, I will do the same for Ethics *in* AI, and Ethics *of* AI respectively. In Section 5, I will show how this taxonomy can answer questions 1, 2, and 3, and will discuss two possible limitations of my analysis.

### 1.1 The Capability Approach

The capability approach (CA) is an approach to compare quality-of-life assessments. It is conceived as an alternative to more consequentialist approaches to measure quality of life. CA is used to assess policies and social arrangements in a variety of contexts, across high-, middle-, and low-income countries. There are a number of formulations of CA (Robeyns 2005), emphasizing different disciplinary angles (such as economics, politics, social sciences, or philosophy), and it is known under many names (e.g., 'human development approach'). I especially rely on Nussbaum's formulation (2011).

[3] Here is the table: https://docs.google.com/spreadsheets/d/1l-t6z1lH5Rk0gTKSiZZbOGjcMMW0uUbq/edit?rtpof=true&gid=1767920905#gid=1767920905

The central point of CA is that individuals not only should have access to concrete positive resources for improving their quality of life, but they also should be able to choose which resources to use, how, and to what purpose. This consideration has a number of ramifications, the most notable being that our assessment of policies or institutions should be concerned with what people are (freely) able to do and be. From these basic considerations, there is a fundamental distinction, which is the one between functionings and capabilities.

Functionings are things that one might value doing or being. Functionings include rather different things, from basic states such as being nourished, to complex activities such as participating to political demonstrations. Most consequentialist theories of well-being tend to focus only on functionings when measuring quality of life, though there is significant variation as to what counts as important functionings for measuring quality of life. CA proposes something different: we should consider not only functionings, but also the freedom that an individual has in deciding which functionings to pursue, how, and why.

These 'freedoms' are called 'capabilities', defined as a range of potential functionings that are concretely feasible for an individual to achieve, and that such individual can freely choose to pursue. The emphasis is on choosing freely, and as such capabilities are seen as *substantial freedoms*, where individuals "may or may not exercise [the freedom] in action; the choice is theirs" (Nussbaum 2011, p 18). These substantial freedoms have an ethical dimension, because (at least in Nussbaum's view) they provide a foundation for human dignity.

There are other important aspects of capabilities, especially concerned to their structure. However, this succinct introduction is enough for the moment, and other aspects of CA will be specified when they become relevant for the discussion of the three kinds of AI ethics.

**1.2 Terminological Caveats**

Let me now turn to a few terminological caveats. In the rest of this article, AI systems will overlap significantly (*but not completely*) with Machine Learning (ML) systems.

I take AI systems to be constituted by two kinds of characteristics: functional and structural characteristics. This distinction comes mainly from philosophy of technology, where a debate stemmed from Kroes' seminal characterization of technical artefacts as having a 'dual nature' (2002). With this, he means that there are two ways of looking at technical artefacts, one *qua* physical systems with certain *structural characteristics*, and one from the point of view of the *goals* they contribute to or task they fulfil. In line with this, admittedly, crude characterization of the complex thesis of Kroes, I take 'functional characteristics' of AI systems as a specification of their goals and tasks, and the structural characteristics as the ones referring to the way AI systems are constituted, both in terms of components, and the procedures followed to coordinate those components[4]. These components belong to ML systems. This means that AI systems are ML systems *plus* something else that goes beyond the boundaries of ML systems themselves, namely tasks and goals (Figure 1).

[4] Kroes' account stresses the importance of the 'material' nature of structural components of technical artefacts. In the case of AI systems, 'structural components' are either virtual (e.g. data sets, algorithms, technical requirements), material (e.g. the hardware that implements the software on which AI is implemented in turn) or hybrids. In this context, I focus especially on 'virtual components', for reasons that will become clear later on

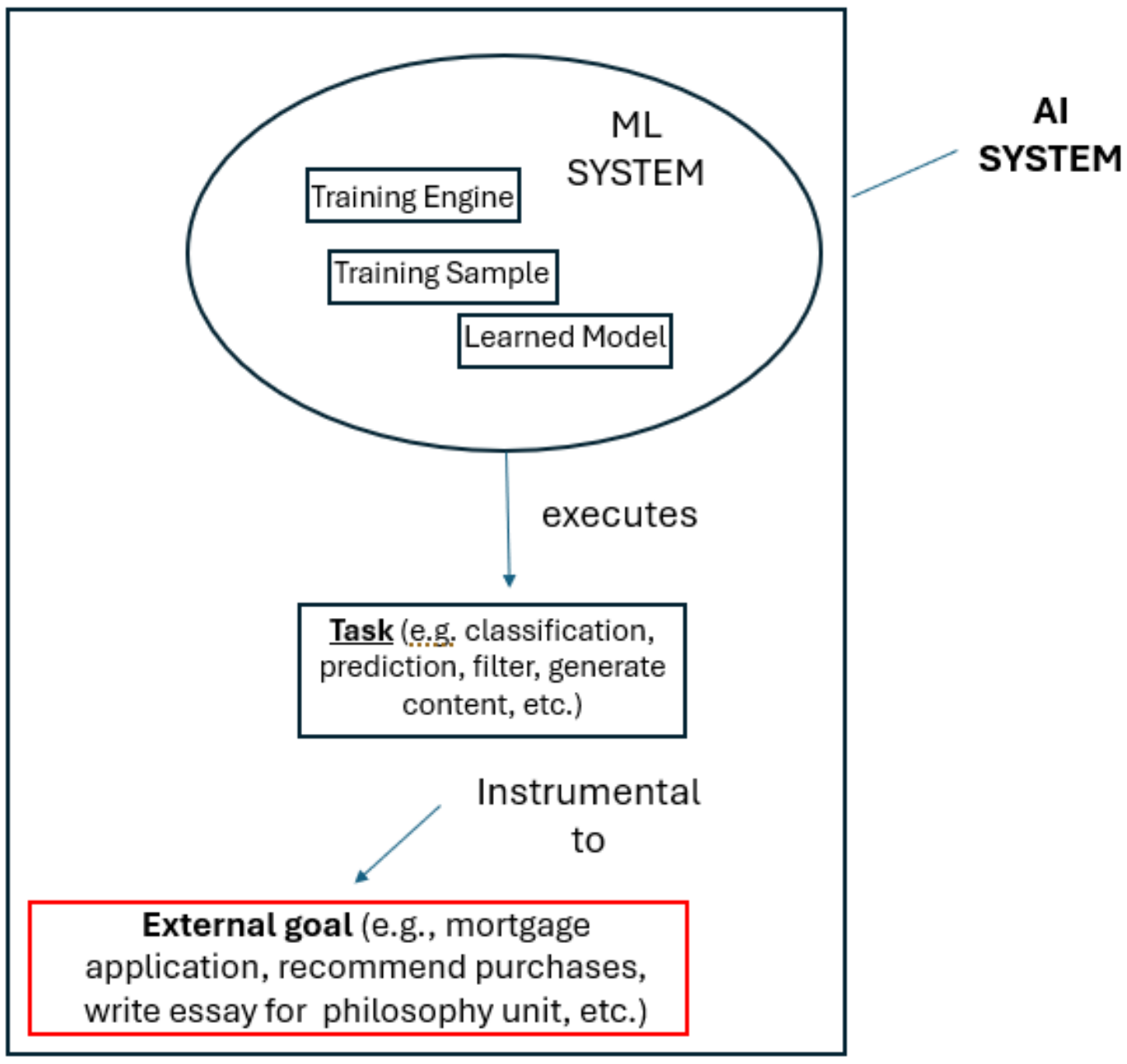

Figure 1. Functional and structural characteristics of AI systems

Structurally, I see AI systems as constituted at least by three components which make up ML systems (Termine et al 2024). First, there is the training sample, which is the repository of data points that the system use to learn and adapt. Second, there is the training engine, which is the computational/optimization 'machine' that the system uses to learn and adapt on the basis of the training sample. Finally, there is the learned model, which is what the ML system 'learns' and it is 'applied' to execute a number of tasks. I include in the category of 'structural characteristics' also the attributes of the procedures and standards followed by AI practitioners to coordinate these components.

There can be a number of functional characteristics or attributes, which falls into two categories (Figure 1). First, there are 'task-attributes'. AI systems can execute tasks such as classification, prediction, content filtering, content generation, etc – task attributes are the 'bare' outputs of an AI system. But these tasks are often instrumental to a goal, which is generally structurally external to ML systems, such as approving mortgage application, recommend purchases, etc. Goals are the 'effects' in the real-world to which bare outputs contribute to. I include also goal-attributes (or 'contribution to goals') as a functional characteristic[5]. For instance, an AI system can classify a user profile (i.e. the task), where the classification is leveraged to process a mortgage application (i.e. the goal). An alternative example is an AI system that filters content (i.e. the task) for a system which then recommends purchases to users (i.e. the goal). In the case of cutting-edge Generative AI, an AI system can generate written content (i.e. the task), that a student will use to write an essay for my philosophy unit (i.e. the goal). Sometimes, functional characteristics are connected to macroproperties of AI systems that might refer to structural components of AI systems, but the technical dimension is overlooked or ignored, as I will show.

Let me now turn to the three kinds of AI ethics.

[5] One could specify more precisely the distinction between tasks and goals by resorting to the distinction between effect role functions and purpose role functions (van Eck 2015). However, adapting this distinction to this context will require more work than it seems, and given the limitation of space I plan to do this in another work

## 2. ETHICS AND AI

The first kind of AI Ethics is what I call Ethics *and* AI. I define it as follows:

Ethics *and* AI (*EaA*) = the study of the relation between the *functional characteristics* of AI systems and pre-existing normative commitments external to AI systems

Central to the definition of EaA are the functional characteristics of AI systems. In the context of EaA, AI systems are looked especially from this point of view. For instance, AI systems can be used to automate mortgage applications, but in doing so they might exclude individuals on unlawful or ethically problematic basis. In this case the problem is with the 'task attribute', namely the 'bare output' of the AI system. But in other cases it is the goal associated to the AI system that is problematic. The EU AI Act (Madiega 2024) prohibits AI systems whose goals are associated with unacceptable risks (such as AI systems used for face recognition tasks). It is also important to point out that the notion of 'unacceptable task or goal' has to be understood from the standpoint of specific normative frameworks (e.g., fundamental human rights). To use an expression from Kaplan's work, EaA is concerned with questions of ethics that are *external* to AI systems, and analogously to what he calls 'philosophy and technology', the goal of EaA is to "analyze technology in terms of ready-made philosophical concepts, usually moral and political concepts such as 'freedom', 'general welfare', and 'human nature'" (2009, p xiv).

In EaA, AI systems are significantly black-boxed from a structural point of view. With this I mean that the attention to the structural characteristics of AI systems is *minimal*. While many examples of EaA stem from considering the opaque nature of ML systems, or the fact that data can be biased, the specific technical facets of these problems are not of particular interests, nor are technical solutions. Rather, what is of central interest to EaA is whether certain outputs or goals of the systems are in tension with specific normative commitments. EaA ethicists' goal is to point out that there is a tension, suggest that the gap between the AI systems and the normative frameworks ought to be filled, and provide in some cases a general and high-level strategy on how to do this (e.g. design publicity of Loi et al 2021). However, how to concretely (viz. technically) do this is a task requiring another skillset and a different set of considerations. This means that EaA ethicists need not considerable technical expertise about AI systems; rather, what is important is that outputs and goals of AI systems may raise some issues of, e.g., privacy or fairness. This is certainly not an easy task, especially when the effort is on showing why certain tasks (e.g. predicting behavior; filtering and ranking information, etc) coupled with macroproperties of AI systems (e.g. opacity, biased data, automated processes) pose specific normative problems, as in the case of showing why trust is indeed an important issue when AI is concerned (von Eschenbach 2021), that automated decision-making systems violate duties of consideration (Grant et al 2023), or that we have indeed a right to an explanation that opacity violates (Vredenburgh 2022). But none of these analyses require any specific technical expertise concerning either opacity or algorithmic bias.

A classic example of EaA is the one based on principles (Jobin et al 2019). The strategy is to take general principles that are seen as constituting the (ethical, political, and civic) backbone of our society, and see to what extent the goals and tasks of AI systems put them in jeopardy. This literature in EaA has taken inspiration from principlist biomedical ethics, often explicitly (Mittelstadt 2019). This move has been misunderstood in various ways - I will come back to the misunderstandings in Section 5. For the time being, consider that the moves licensed by principlism in AI Ethics are not based on dubious similarities between AI and biomedicine as a profession (as correctly criticized by Mittelstadt 2019). Rather, they are based (or should be based) on the idea that the principles of biomedical ethics are universal principles, at least in Beauchamp and Childress' perspective (2009). In particular, the source of the four principles of biomedical ethics is what they call 'common morality', which "is applicable to all persons in all places, and all human conduct is rightly judged by its standards" (Beauchamp 2007, p 7). But developing AI systems is one instance of 'human conduct', and hence it has to be judged by the standards of those principles. This means that the 'principles' of EaA are conceptualisations of normative concerns and requirements that, at least from this perspective, stand for general desiderata of our society, and that everyone will likely accept (at least from the principlist perspective). The only genuine AI-based principle seems to be explainability, but it is not difficult to show that explainability can be reduced to a combination of other more general principles, such as justice and respect for autonomy, the importance of trust (von Eschenbach 2021), or in general that 'explaining' can be reconceptualized as 'justifying' AI systems in the face of the normatively-laden goals (Loi et al 2021). In principlist EaA, the goal of AI ethicists is to gauge the feasibility and appropriateness of AI systems and the goals they serve within a given context. AI ethicists will orient discussions about those normative commitments in the context of AI systems, to make sure that high-level principles are fruitfully interpreted and/or developed. Textbook examples of EaA based on principles are not difficult to find. For instance, Char et al (2020) looks at the classic ethical issues raised in biomedicine, and then adds 'AI' to the healthcare context to see how this complicates the situation – which, according to the authors, it does. A rising trend in EaA is to pay attention to the principle of 'safety', which can be redefined in terms of 'non-maleficence' (Gyevnar and Kasirzadeh 2025).

While EaA is dominated by principlism, there are also cases of principle-free EaA. For instance, Waeken (2022) redefines ethical issues typically framed in the principlist way in terms of *power*. The central concern is about emancipation and empowerment, and because there is evidence that goals associated to AI systems might jeopardize them, then AI Ethics should be framed in terms related to critical theory.

CA is also well-represented in EaA, and it is another example of principle-free EaA. In 2019, a roadmap for supporting a more equitable development of AI was published by the United Nations (2019), in line with a plan for achieving the so-called Sustainable Development Goals. This was part of a larger effort by United Nations to address the emerging political and ethical issues raised by frontier technologies. AI here is not considered a unique technology: the main issue is what risks cutting-edge technologies like AI raise for Sustainable Development Goals in general. What emerged from this document is that AI systems should be evaluated on the basis of how well they align with already pre-existing developmental goals and human rights, and that whatever AI system one wishes to develop, it should "balance economic, social, and environmental goals" (p 3), where those goals are understood on the basis of the categories provided by the United Nations' frameworks. In line with the Human Development Approach (namely, CA) endorsed by United Nations, the emphasis and the focus is especially on low- and middle-income countries, which are typically at the mercy of high-income countries when emerging technologies (and an equitable share of burdens and benefits) are concerned. More specifically, four distinct layers of 'capacity development' are identified for AI, from infrastructure, to data, human capabilities, and human rights-based laws and policies. All those typical issues raised by a CA-based approach to policy-making are present here: issues of digital divide, conversion factors related to capability-expansion, and focus on human rights. Therefore, the goal of CA-based EaA, as envisioned by United Nations, is to direct all AI-related policies towards those concerns typically raised by CA. EaA will then shape the AI community by directing the attention of their practitioners towards those CA-based concerns

## 3. ETHICS IN AI

A second kind of AI Ethics that is well-represented in the literature is what I call Ethics *in* AI, which is defined as follows:

Ethics *in* AI (EiA)= the study of how AI systems, given their design flexibility, can be constructed such that their *structural characteristics* reflect given ethical and/or political commitments

In line with Section 1, by 'structural characteristics' I mean the internal characteristics and components of ML systems, in particular the training engine, the training sample, and the model which is constructed as a result of the training of the engine on a given data set (Termine et al 2024). Each of these components is constructed and coordinated by data scientists/AI practitioners by following certain procedures on the basis of benchmarks and standards. As I noted above, I include those procedures and their standards in my definition of 'structural characteristics'.

EiA is based on the idea that ethical, societal, and political issues raised by AI systems are often failures of design, in the sense of failures to pick up and coordinate the right structural characteristics. AI systems generate outputs that are unfair, violate privacy, or jeopardize safety, because they have been designed to generate those outputs. If this is true, then it is possible to design AI systems such that they will deliver the right ethical outcomes, and this can be done by modifying the structural characteristics of AI systems themselves.

These ideas have emerged in an engineering context that sees ethics as a matter of 'techno-fixes'. Ethics is not about a community effort to orient functional characteristics of AI systems towards the right normative desiderata. Rather, ethical problems are just one set of problems that could be addressed with an engineering mindset. As noticed by Wiggins and Jones (2023), most of 'techno-fixes' have focused especially on privacy and fairness, leading to a proliferation of attempts to 'code' fixes into AI systems – e.g., k-anonymity, differential privacy, independence, separation, sufficiency, etc. Especially when it comes to fairness, there has been also an explosion of workshops and meetings dedicated explicitly to the construction of novel techno-fixes (e.g., FAccT).

Unlike EaA where AI systems are black-boxed, in EiA AI systems are indeed opened up, and their structural characteristics shaped to make sure that they deliver the right results. Special attention is especially devoted to output metrics and data preparation, which is where concerns of privacy and fairness are more likely to emerge. This emphasis is structural rather than functional: it is not necessarily about the outputs of algorithms; rather, it is about what kind of metrics we use for measuring the ethical relevance of outputs, whether one metric rather than another reflects our moral and political commitments, and once we pick up the right measure how we modify the system in such a way that the outputs will be 'ethically correct'. This has been explicitly raised in the debate between ProPublica and Northpointe/Equivant on the alleged discriminatory nature of COMPAS, where one side accused the other of using the wrong notion of fairness to inform choices regarding how to measure 'discrimination' (Ratti and Russo 2024): while Northpointe/Equivant (on the basis of the metric that they used, called 'calibration') claimed that they need not take any action for modifying the AI system because the outputs were not discriminatory, ProPublica (on the basis of the metric that they considered relevant, namely predictive equality) argued otherwise (Castro 2022).

The contribution of ethics as a discipline to EiA is to provide the right conceptual resources related to those moral and political commitments that the structural characteristics of AI systems ought to reflect. In the literature on techno-fixes, this has been done especially by data scientists or computer scientists: an EiA ethicist is often a well-rounded computer scientist or data scientist who takes normative concerns as something internal to, or part and parcel of designing AI systems.

Representative of these efforts is the rich literature on *value alignment*. A classic of this approach is Gabriel's work (2020), which considers the different philosophically-informed standpoints through which aligning AI to human beings' normative commitments or, briefly put, to human values. Moreover, Gabriel reviews the main challenges for this endeavour, from formally encoding values or principles in AI systems, to choosing the right principles on the basis of resources provided by philosophy, social sciences, and political theory. But computer scientists or data scientists need not be AI ethicists. In fact, philosophers, social scientists, and political theorists have been collaborating with computer and data scientists exactly to contribute to the process of choosing the right ethical and political conceptual tools, as well as implementing them. For instance, in a seminal article Binns (2018) shows how works in moral and political philosophy can inform debates about fairness in ML.

It is useful to illustrate an example of EiA that uses CA as a resource to design AI systems that deliver the correct outputs. London and Heidari (2024) are concerned that most approaches to value alignment are too centrally focused on the values stemming from the cultural background of creators (that is, high-income countries), and that they are not emphasizing larger impacts of AI. The importance of interacting with AI systems in such a way that AI systems are meaningfully beneficial to individuals, they say, cannot be adequately prioritized in current AI alignment strategies. Doing this requires reconceptualizing AI systems as *assistive technologies* – which means redefining their functional attributes (that is, the goals and functions that AI systems serve). But in order to do this, it is necessary to lay out precise structural characteristics of AI systems as well, which they do in terms of a formal characterization of Nussbaum and Sen's CA. This is because, in their opinion, redefining the structural characteristics on the basis of CA is the only way to build AI systems that are indeed assistive technologies. This is a convincing case of EiA: AI systems are conceptualized and structurally characterized through a theoretical framework provided by economics and philosophy.

## 4. ETHICS OF AI

The third kind of AI Ethics is what I call Ethics *of* AI. This is defined as follows:

Ethics *of* AI (EoA) = the study of how AI systems and the communal practices of the contexts in which they are implemented, shape each other

Central to EoA is the idea of 'communal practice'. This term has its origin in the context of the contemporary revival of virtue ethics, especially the one inspired by MacIntyre's work (2011). What a practice is in this context is a vexed question (Dunne 1997), and I do not have the space to cover such a large topic. I will start by MacIntyre's definition and then draw a number of limited considerations. 'Communal practice' is defined as

> "any coherent and complex form of socially established cooperative human activity through which goods internal to that form of activity are realized in the course of trying to achieve those standards of excellences which are appropriate to (…) that form of activity, with the result that (…) human conceptions of the ends and goods involved, are systematically extended" (p 218)

There is a lot to unpack in this definition. 'Complex' refers to the idea that communal practices can be described at a number of different levels, while the collaborative nature of communal practices refers to the fact that they require individuals to negotiate their goods, standards, and procedures. But most important for EoA, communal practices are sustained activities that are goal-oriented, and as such require certain normative conceptions of what count as a good for that practice, and which are the legitimate means to get to those ends. Examples of communal practices include science, higher education, medicine, football, etc. What characterize these communal practices are that, in addition to instrumental goals, they also aim at intrinsically valuable goals (e.g., scientific knowledge, educating students, curing patients, etc), and they also specify legitimate means to achieve those goals (e.g., the scientific method and scientific integrity, principles of biomedical ethics, etc).

How is 'communal practice' connected to ethics? Common examples of communal practices – science, medicine, theatre, education, chess, football, etc – show the importance of a collective effort in which individuals might compete with each other, but also work together "to advance shared goods seen as having intrinsic values" (Hicks and Stapleford 2016, p 454). As such, it is possible to see communal practices as those spaces where individuals, in collaboration with other individuals, act and make choices that are instrumental to pursue goals that are considered intrinsically valuable and, in some cases, are relevant to live the kind of life that individuals have reasons to value. This is especially noteworthy in communal practices with an essential aspirational character, such as science (Ratti and Stapleford 2021), or higher education. Being these spaces where individuals pursue their life aspirations, communal practices are tightly connected to implicit or explicit conceptions of the Good Life. But ethics is, by definition, about choices and actions as they unfold with respect to conceptions of how we ought to live. Therefore, the structure and the dynamics of communal practices are ethically salient, because they will have effects on questions about agency and choices which are relevant for the Good Life.

What is the relation between this and AI Ethics? Given the connection between communal practices and ethics, AI systems can be said to 'mediate' the Good Life, because they 'mediate' our experiences in the environments and the contexts in which communal practices unfold. Here I take 'mediation' in the technical way this term has been understood in philosophy of technology (Ihde 1990; Verbeek 2004). At a very basic level, AI systems 'mediate' as any other technical artefacts do, namely by shaping our perceptual and interpretative abilities[6]. The specific way AI systems do this is typically by shaping, constraining, and transforming the same environment in which we act and make choices (Danaher 2016). While this applies to other technical artefacts as well, the scale (Creel and Hellman, 2022) and the invisibility (Moor 1985) of AI systems is unprecedented. AI systems can mediate the Good Life because, by shaping and constraining environments in such radical ways, they can also transform our communal practices.

Consider a general example now, which will be used to build on more specific ones later. Take the online world – a significant part of our existence is spent online. Living online shape the way we pursue our lifegoals, as we continuously interact with individuals with different viewpoints, and we are exposed to an increasing amount of information regarding goods that can be relevant to our lifegoals and their means. And now consider how recommender systems shape that environment (Milano and Prunkl 2024), without us even noticing it – often, what happens is that these AI systems can insulate us in nefarious ways, and deprive users of alternative conceptions of goods, ends, and life aspirations, thereby profoundly shaping communal practices. But what is especially noteworthy is that it is difficult to anticipate these effects on the basis of functional and structural characteristics of AI systems *taken in isolation*. Users with opportunities to learn and conceptualize life aspirations outside the online environment will not be constrained by an echo-chamber, while others will be (Vallor 2016). In order to properly anticipate and gauge effects on communal practices and the Good Life, one needs to understand how the structural and functional characteristics of AI systems 'interact' with the context in which AI systems are used. In other words, the influence of AI systems on our environment – and hence possibly our communal practice - come from how structural and functional characteristics

[6] There is a lot to say about this, but for reasons of space, consider Ihde's (1990) and Verbeek's (2004) works

‘react’ to the characteristics of the context of implementation, and this can happen in rather unpredictable ways.

EoA is exactly interested in discerning the relationships between functional and structural characteristics of AI systems on the one hand, and the context in which AI systems are implemented on the other, where the goal is to uncover possible effects on communal practices. Because of this interest, EoA is a genuine sociotechnical approach (Fazelpour and Danks 2021): it focuses on structural and functional characteristics of AI systems (i.e., the ‘technical’) and the characteristics of the context in which such systems are employed (i.e., the ‘socio’). The goal of EoA ethicists follows from the above considerations. Given an AI system *x,* with functional and structural characteristics *c1, c2, …, c3* to be implemented in an environment *y* with features *f1, f2, …, fn* , the goal of an AI ethicist is to analyze how *cs* might negatively shape communal practices in *y*, where this analysis is based on a thorough investigation of *cs* and *fs*. This analysis is typically carried out from the standpoint of a framework coming from a number of disciplines, be they moral philosophy, political theory, social sciences, etc.

Let me now turn this picture into something more concrete by illustrating EoA through CA. In the case of EaA, CA is used as an external framework through which evaluating the functional characteristics of AI systems. In the case of EiA, CA is used to evaluate the structural characteristics of AI systems. In the case of EoA, CA is used as a method to make sense of the relation between AI systems and communal practices (Ratti and Graves 2025). As I have explained in 1.1, capabilities are functionings that are concretely achievable by individuals, and that individuals can freely choose to pursue. Nussbaum (2011, pp 33-34) identifies ten central capabilities, which are deemed relevant for human dignity, and the task of a government is to ensure that all citizens possess at least a threshold of these capabilities. Because capabilities are about choices, actions, and agency, if AI systems mediate actions by restructuring the environment where communal practices unfold, then they shape capabilities. But this does not say much: it is just a different formulation of the basic idea described above, namely that AI systems shapes our choices (i.e., capabilities) by structuring the context where communal practices unfold.

CA becomes helpful only when we pay more attention to the structure of capabilities, which Nussbaum calls *combined capabilities*. In addition to capabilities proper, combined capabilities include *conversion factors*. Whether a functioning is effectively achievable – whether it is indeed a capability – will depend on factors that allow a person to freely turn (viz. convert) a possibility into something actual. Therefore, whether one can make a choice that matters to them, will depend especially on these conversion factors. There are several types of conversion factors, including personal (e.g. reading skills, physical condition, metabolism, etc), social (e.g., public policies, social norms, societal hierarchies, etc), environmental (e.g. climate, geographical location), and digital (e.g. access to computers, phone network, broadband). In order to understand the connection between capabilities and conversion factors and how they generate combined capabilities, take a classic example of the CA literature. Bicycles are technical artefacts that can potentially expand a number of central capabilities. For instance, the capability of affiliation can be greatly expanded as a result of, say, young kids using bicycles to join a sport club in a nearby town. Senses, imagination, and thought can be also expanded as a result of individuals biking outside large urban areas and engage in nature sightseeing. However, the functional and structural characteristics of bicycles *alone* cannot guarantee that capabilities will be expanded, nor we can predict that they will do only on their basis. In order for bicycles to do so, several conditions must apply: one needs to have the right physical factors enabling the use of bicycles; a certain infrastructure facilitating the movement of bicycles must exist, etc. For instance, a bicycle in the Netherlands (with its extensive bike path) will expand capabilities better than in the Amazon

Forest. In other words, bicycles can convert possibilities into actual functionings, only if some personal, social, and environmental factors are already in place (that is, if conversion factors are present).

AI systems are no exceptions. Consider a famous case in medical AI discussed a few years ago. Obermeyer et al (2019) investigate the performance of a widely used health-risk algorithm. The goal of this AI system can be conceptualized as expanding health as a capability (namely, pursuing health-related goals that one can freely choose to pursue), where this capability is a necessary condition for engaging in communal practices in general[7]. What this example shows is that a lack of attention to the factors characterizing the context of implementation can result in AI systems causing a lot of damage of ethical relevance. In their study, Obermeyer et al found that the AI system falsely concluded that Black patients were healthier than equally sick White patients, though the algorithm appeared to be well calibrated across races. What emerged from their analysis was that the system used 'health expenditure' as a proxy for health risk. It is not unreasonable to think about 'health expenditure' as a proxy for risk: the more one spends on health, the more this person might have health-related problems. However, 'health expenditure' is a proxy only for those individuals who already have access to healthcare, where 'health care access' is a conversion factor (Prah Ruger 2010) that depends in turn on other conversion factors, such as having a full-time job; health insurance; a stable income; living in an area where healthcare is accessible; etc. In other words, the AI system was performing well only for those individuals living in a context characterized by a number of factors of a specific personal, social, and environmental nature. AI practitioners have assumed that all end-users had a homogeneous level of conversion factors; however, with this controversial assumption, they automatically exclude and make invisible all those who lack that particular level of conversion factors. This means that structural characteristics of AI systems (i.e. the ones determining 'health care access' functioning as a proxy for health risk) have interacted with the characteristics of the context of implementation (i.e. the lack of certain personal, social, and environmental factors characterizing the context of certain end-users), in such a way that the AI system would make some end-users invisible to the radar of healthcare assessment risk, thereby potentially affecting their participation to communal practices. But it is difficult to anticipate this unfortunate outcome on the basis of structural and functional characteristics of AI systems *alone*. What the CA-based EoA suggests instead is that one ought to look at the AI system *and* the context of implementation, by identifying the relations between functional and structural characteristics on the one hand, and conversion factors on the other.

There are a number of interesting works in EoA in the literature. Some works are devoted to make broader points. For instance, Heuser et al (2025) argue in favour of a more comprehensive AI ethics, which also includes an analysis of how AI systems and the lifeworld interact sometimes with surprising outcomes, where the idea of lifeworld is connected to ethics in a way comparable to how I have connected communal practices and ethics above. Other contributions take a particular perspective on the relation between AI systems and communal practices from a more explicit normative standpoint. For instance, Longo (2025) describes how algorithmic systems embedded in social media mediate political judgement. Most important, he provides an analysis of how AI does not 'replace' or 'determine' judgement as others have argued; rather, AI systems mediate (in a postphenomenological sense) judgement. This analysis is done from the standpoint of Arendt's normative views (in the sense that Arendt's perspective is helpful for illuminating these dynamics).

---

[7] In fact, some think (Nussbaum 2011; Venkatapuram 2011) that 'health' is not a proper capability; rather, it is a functioning that can be characterized as a necessary condition for all other capabilities to be expanded. The same can be said for communal practices: without health, it is difficult to engage in any meaningful communal practice

## 5. IMPLICATIONS

To summarize, Figure 2 visualizes the relation between ‘ethics’ and ‘AI’ in the three kinds of AI Ethics. In the case of EaA, AI systems and ‘ethics’ are not overlapping. There is an arrow going from the ‘ethics’ to the AI system, which represents the idea that a normative framework is brought to evaluate the functional dimension of AI systems. This emphasizes the importance of a community effort to bring AI and ‘ethics’ closer by making functional characteristics of AI systems compatible with ‘ethics’. In the case of EiA, ‘ethics’ is literally inside AI systems: it is conceptualized as part and parcel of the structural characteristics of AI systems. Finally, in EoA, because the focus is on the mutual relationships between AI systems and the context of implementation from the perspective of a given normative framework, the system and the context are ‘contained’ in the ’ethics’.

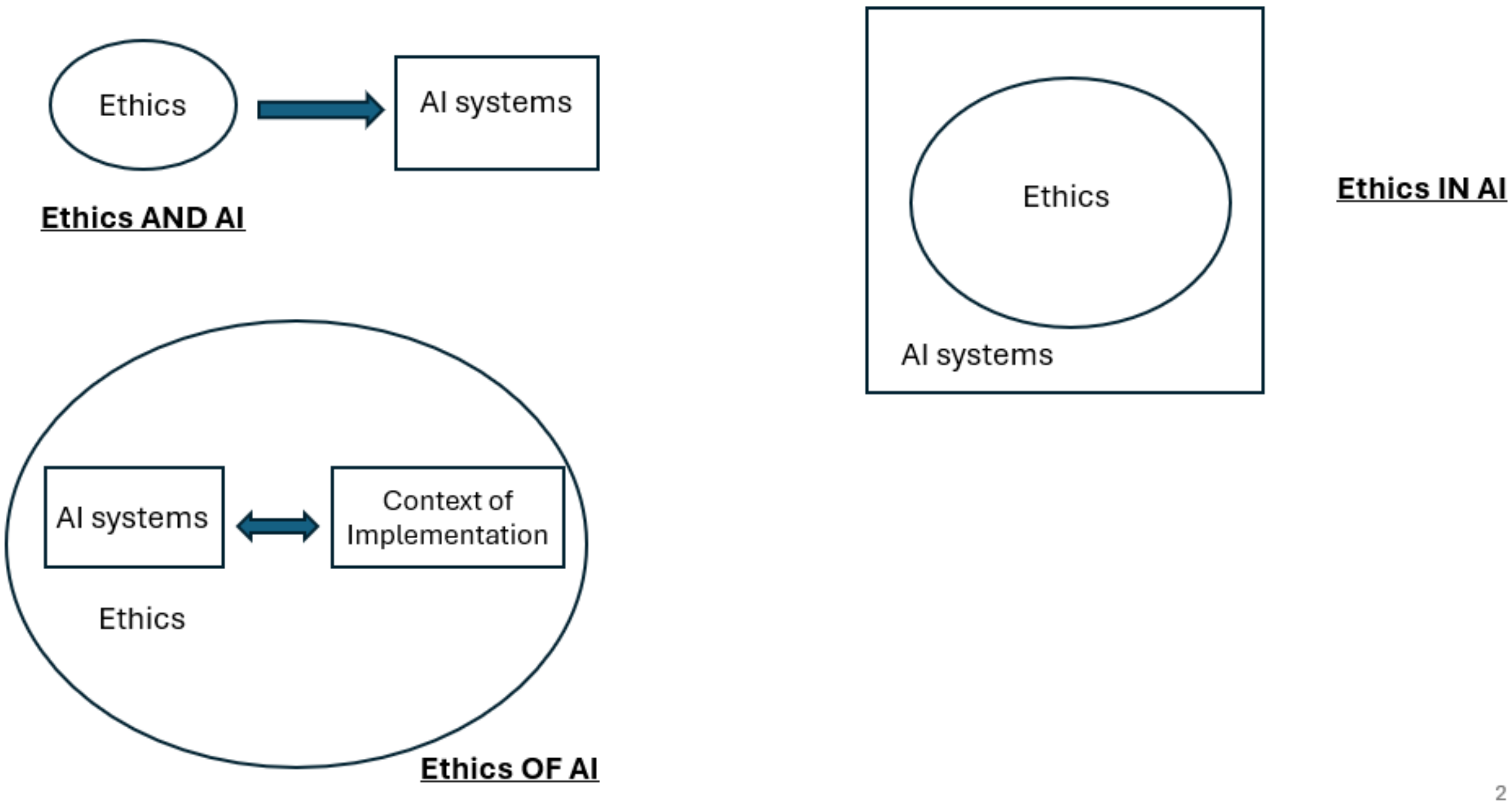

Figure 2. Relationships between ‘ethics’ and ‘AI systems’

This characterization of AI Ethics provides a vantage point to answer questions 1 and 2 raised in the Introduction. Let me start with 1. Questions that AI ethics aim to answer will differ depending on the kind of AI ethics considered. For instance, in the context of EaA, questions concern the normative implications raised by AI systems, considered at the functional level: does opacity of AI system undermine public trust? Do we have a right to an explanation to how outputs have been delivered by AI systems? Who is accountable for the outputs of automated AI systems? In the case of EiA, questions concern the actual implementation of ethical and political concept into AI systems: what kind of privacy concepts can we implement in AI systems? How should we handle impossibility theorems? What is the best way to implement ethics in the corporate pipeline? In the case of EoA, questions will be about the mutually shaping interactions between AI systems and communal practices: how do AI systems affect the doctor-patient relation? How do generative AI systems shape higher education and the cultivation of specific intellectual skills?

Let me now turn to question 2. Given the specific questions each kind of AI ethics aims at answering, different expertise and skillsets are needed. In order to explain the different kinds of expertise at play, I take Evans and Collins’ (2002) famous distinction between *contributory* and *interactional expertise*. While the context in which they introduce this distinction is markedly different than the present one, their ideas are nonetheless relevant for this article. On the one hand, contributory expertise in a field *x* refers to the kind of expertise that one needs in order to provide a valuable contribution within *x*, say publishing an article in a specialized peer-reviewed journal, or designing a discipline-specific methodology. On the other hand, interactional expertise in a field *x* designates the level of expertise that a practitioner of a field *y* needs to interact and work successfully with practitioners of the field *x*. For instance, there are philosophers of biology who regularly works with biologists, and while they might not have a level of expertise that will allow them to, e.g., design and perform experiments, they have enough knowledge of biology to have meaningful conversations with contributory experts of the biological sciences. In the case of the questions pertaining EaA that I mentioned above, contributory expertise will cover law, moral and political philosophy, and political theory, while a minimal and basic level of interactional expertise in AI is needed. In the case of EiA, a contributory expertise in AI and organization/management science is needed, complemented with an interactional expertise in ethics and political philosophy. In the case of EoA, a contributory expertise in applied ethics, philosophy of technology, and the social sciences (especially to find ways to 'measure' the mutual influence of AI systems and communal practice) is required. An interactional expertise in AI is all you need, even though at a more sophisticated level than the one required in EaA, given that an EoA ethicist needs to be familiar with the structural characteristics of AI systems. This is because, what is important is how AI systems – covering both functional and structural characteristics – shape communal practices..

Having answered 1 and 2, let me now turn to 3. My account reveals some limitations of classic criticisms against AI Ethics. Consider the harsh criticisms raised against the approach in EaA based on principlism. For instance, Munn (2023) provides a list of flaws of principle-based AI Ethics. Principles are deemed meaningless, in the sense of being "highly abstract and ambiguous, becoming incoherent" (p 870), and a gap with practical implementation has been also considered as a fatal flaw for this kind of AI ethics (Morley et al 2020). Other problems include that lack of consequences resulting for not complying with principles. These are important concerns, but it is important to realize that, from a principlist and EaA perspective, they are simply misplaced. First, the generality of principles makes more sense when contextualized within the historical emergence of principlism in applied ethics (Wiggins and Jones 2023). Arguably, its origin is the *Belmont Report* in 1978, which set up principles and guidelines for research on human subjects after years of discussions. The report conceptualized ethics as a negotiation of tensions concerning means and ends, on the basis of the acceptance of three principles that were seen as 'epistemic backstop', "the consensus on which all parties can agree, even when disagreeing about specific applications" (Wiggins and Jones 2023, p 239). Principles in the Report are explicitly 'comprehensive' and general enough to cover present and future normative concerns. This is related to a second (misplaced) criticism: how can these general (and meaningless, toothless, remote from practice, to use Munn's terminology) principles apply? This problem of 'applicability' can be overcome by considering the nature of principles as such. Principles are not rules, as rules trump further reasoning. Unlike rules, principles allow for flexibility (Ratti and Graves 2021). This means that one does not follow principles; rather, one *weighs* principles. Therefore, if we say that we cannot apply principles because they do not provide enough information to be followed, then we are simply ascribing to principles a task that should be assigned to rules. But what does it mean to weigh principles? We can think about 'principles' in applied ethics as akin to what is prescribed by constitutions, where these provide orientation of a general nature, and it is then up to communities to formulate more specific standards for individual cases. We should not expect principles to do all the work, as rules do; rather, it is an individual community "which does the hard work to distil these principles into standards, rules, and therefore into practice" (Wiggins and Jones 2023, p 240).

Therefore, some of the criticisms of the dominant form of EaA (i.e. principle-based) simply miss what principlism is up to. And most important, this idea that normative commitments like principles are just orienting design and implementation of AI systems within a specific context, can be used to anticipate similar objections of limited applicability that can be raised to other forms of EaA: the goal of AI ethicists, as shown above, is to facilitate discussion leading to the formation of the backbone of these communities, and it is up to these communities to distil the general normative commitments into standards and benchmarks. This suggests that harsh criticisms against EaA have been made from the perspective of EiA: if a proposal in EiA had the same characteristics of a principlist-based EaA, then criticisms made by, e.g., Munn, would be on target. However, principlist-based AI Ethics is often EaA, so those criticisms are unfair. Similarly, it has been sometimes said that off-the-shelf tools provided by EiA tend to uncritically assume certain underlying normative concepts, where choice between them is value-laden and requires "philosophical arguments and considerations that fall outside of the narrow technical scope of the standard approach to fair ML" (Fazelpour and Danks 2021, p 10). But one can argue that this is more the role that EaA ethicists should play, rather than EiA ethicists. In other words, EiA ethicists need not to provide arguments as to why certain normative commitments are important in the case of AI – this is a task for EaA ethicists.

These last considerations raise the question about the relation between the three kinds of AI ethics. Are these approaches in competition one with the other? Are they complementary? I do not have a definitive answer to these questions. At first glance, there seems to be a relation of continuity between EaA and EiA. Because EaA ethicists orient discussions around AI systems towards relevant normative commitments that are seen as valuable by a given society, then it is reasonable to conclude that they are also moving the first steps towards the process of implementing specific moral and political concepts in AI systems (namely, EiA). In fact, one can see an EaA ethicist playing a role also as an EiA ethicist, by enumerating to computer and data scientists the impressive variety of conceptual tools that can be possibly formalized. But my considerations about the expertise needed in different kinds of AI ethics will block such a move: EaA and EiA require different expertise. Let me stress this with regard to EiA. A significant portion of EiA takes place in private companies. We are all well aware of the highly controversial outcomes of implementing AI ethics in private companies, such as the elimination of the Google's Ethical AI Team which led to the firings of influential EiA ethicists such as Timnit Gebru[8]. As documented by Metclaf et al (2019), it is not just a matter of finding the 'right ethics' to implement in AI systems. To succeed, EiA ethicists need to persuade members of the organization of the importance of the nature of the constrains that EiA can put on AI systems themselves. As Wiggins and Jones point out, "it is unclear how to convince colleagues to value ethical principles [that] could serve as any constraint – particularly if those constraints would reduce profit" (p 245). This is to say that knowledge and experience of corporate and management dynamics, which is not required to EaA ethicists, is an essential skill for EiA ethicists. But one might also say that this relation of continuity can be characterized as a hierarchical division of labour between EaA and EiA, where the former indeed sets the kind of normative commitments that should be implemented structurally in AI systems. This is an interesting proposal, and I tend to be in favor. EaA ethicists' work is in the context of hard and soft regulations of AI systems, and they provide a normative orientation for how AI systems should be designed, not only functionally, but also structurally – this is because the outputs of AI systems reflect, to a certain extent, their structural constitution. This does not pose any issue regarding the expertise needed: those who are in charge of implementing structurally these commitments will be EiA ethicists – and this reflects the story about principlism and the moot criticism that principles are difficult to apply.

[8] https://www.wired.com/story/google-timnit-gebru-ai-what-really-happened/

Let me now turn to a possible limitation of my account: is there something that has been left out? There is indeed one promising trend in the literature. There are a number of articles (Grosz et al 2019; Bezuidenhout and Ratti 2021; McLennan et al 2022) doing exciting research on how to teach AI ethics to engineering students, researchers, or practitioners. Particularly famous is the so-called 'embedded ethics' approach (Grosz et al 2019), launched by Harvard University[9]. This approach to teaching AI ethics is based on the idea that systematic exposure to ethical problems as they arise in technical contexts, will habituate students or practitioners to anticipate ethical pitfalls, or simply will create a 'feeling' for ethics, that is now missing. As shown in Grosz et al's foundational article (2019), in embedded ethics students learn about the ethical implications of AI "while they are learning ways to develop and implement algorithms" (p 56). As a consequence, ethics modules are not about exposing students or practitioners to moral concepts or theories *in abstracto*, as it is often done in traditional units in moral or political philosophy taught in technical curricula; rather, students will be exposed to morally charged situations in the context of technical units that they are already attending. There are two reasons why I have not created a separated category for this noticeable trend. First, while the number of embedded ethics programs around the world is growing, there is not much writing about it. There are a few attempts at theorizing (Bezuidenhout and Ratti 2021; Ferdman and Ratti 2024), as well as more methodological articles on how to measure impact (Kopec et al 2023). However, much more has to be theorized and conceptualized about embedded ethics, and for this reason an *ad hoc* category might not be warranted. Second, we can conceptualize embedded ethics as an approach to AI ethics education, rather than an approach to AI ethics *as a discipline*. From this perspective, embedded ethics can be tailored around the three kinds of AI ethics, depending on the particular needs of who is involved in the ethical training. In other words, embedded ethics is complementary to AI ethics as a whole, rather than constituting a part of it.

Finally, the reader may have noticed that in Supplementary Table 1 there is a column for 'Academic Ethics and AI'. I think about this as a variety of EaA, because in this literature ethical questions about AI systems are external to AI systems themselves, as in traditional EaA. However, the angle is slightly different. In Academic EaA, the focus is not on how the relationship between AI systems functional characteristics and external normative frameworks. Rather, Academic EaA takes AI as an interesting case study to develop an already existing discussion in an academic debate. In Academic EaA, it is shown that AI can shed light on the debate on a certain concept (or not, contrary to expectations or previous works). For instance, in Schuster and Lazar (2025), judicious attention allocation is the main topic, and AI is treated as an interesting illustration of the normative problems it raises. In other words, there is a pattern of moral problems related to attention allocation that philosophers and social scientists have been discussing for a while, and AI follows (and possibly adds to) this pattern. AI raises problems of attention allocation because of the goals that AI systems are typically built for (e.g. nudging users).

## 6. CONCLUSION

In this article, I have provided a structured analysis of AI ethics as a discipline. I have distinguished three different senses in which AI ethics has been understood so far, highlighting the research questions, role for AI ethicists, problems and prospects for these three kinds of AI ethics. I have also shown the possible relations between the three kinds, and I have highlighted a number of interesting trends in the literature that, in the future, might enrich my analysis. All in all, this article is important to understand that AI ethics is not one thing; rather, it is many things, and there is currently no AI ethicist that can reasonably claim to be able to cover all three sets of questions, methods, expertise, and skillsets. This article will hopefully orient more informed discussions on what the role of AI ethicists is, their limits, the expertise needed, and their training.

[9] https://embeddedethics.seas.harvard.edu/